\documentclass[aps,prd,a4paper,superscriptaddress,nofootinbib,
showpacs,twocolumn,showkeys,amsfonts,amssymb,amsmath]{revtex4}
\usepackage{amssymb,latexsym}
\usepackage{amsmath, amsthm}

\usepackage{amscd}
\usepackage{amsthm}
\usepackage{times}
\usepackage{epsfig}
\usepackage{psfrag}
\usepackage{graphicx}
\usepackage{amssymb,latexsym}

\begin{document}

\title{The instability of black hole formation in gravitational collapse}

\author{Pankaj S. Joshi} \email{psj@tifr.res.in}
\affiliation{Tata Institute of Fundamental Research, Homi Bhabha road, Colaba, Mumbai 400005, India}
\author{Daniele Malafarina} \email{daniele.malafarina@polimi.it}
\affiliation{Tata Institute of Fundamental Research, Homi Bhabha road, Colaba, Mumbai 400005, India}

\swapnumbers
\begin{abstract}
We consider here the classic scenario given by Oppenheimer,
Snyder, and Datt, for the gravitational collapse of a massive matter
cloud, and examine its stability under the introduction of small tangential
stresses. We show, by offering an explicit class of physically valid tangential stress
perturbations, that
an introduction of tangential pressure, however small, can qualitatively
change the final fate of collapse from a black hole final state to a
naked singularity. This shows instability of black hole formation in collapse
and throws important light on the nature of cosmic censorship
hypothesis and its possible formulations. The key effect of these perturbations
is to alter the trapped surface formation pattern within the collapsing cloud
and the apparent horizon structure. This allows the singularity to be
visible, and implications are discussed.

\end{abstract}
\pacs{04.20.Dw,04.20.Jb,04.70 Bw}
\keywords{Gravitational collapse, black holes, naked singularity}
\maketitle


The continual gravitational collapse of a massive
matter cloud within the framework of general relativity was
investigated for the first time by the
classic works of Oppenheimer and Snyder, and Datt (OSD)
\cite{OSD}.
Such a treatment of dynamical collapse would be
essential to determine the final fate of a massive collapsing
star which shrinks catastrophically under the force
of its own gravity when its internal
nuclear fuel is exhausted.

The outcome in the above case is seen to be a
black hole developing in the spacetime. As the gravitational
collapse progresses, an event horizon forms within
the collapsing cloud and from the region
within the horizon no material particles or light rays can escape,
thus forming a black hole. The continually collapsing
star enters the horizon and finally ends up forming a
spacetime singularity, which is hidden inside the black
hole and which is unseen to all the outside observers
in the universe. The matter and energy densities, spacetime
curvatures, and all physical quantities blow up and
take extreme values in the limit of approach to
such a spacetime singularity.

This classic picture became the foundation of
an extensive theory and astrophysical applications
of modern day black hole physics, further to the
suggestion that {\it all} realistic massive stars undergoing a
continual gravitational collapse would have the same
qualitative behaviour. This means that, while the general
theory of relativity necessarily implies the formation
of a spacetime singularity as the endstate for a massive
collapsing star, such a singularity will always be
necessarily hidden within a black hole. Such an assumption
is known as the cosmic censorship hypothesis
\cite{Penrose},
and taking
it to be valid, the theory and applications
of black hole physics have developed extensively in
past many decades.

The cosmic censorship has, however, remained an
unproved conjecture as yet in gravitation theory, despite
numerous attempts to establish the same.
Therefore, in past many years, much effort
has also been devoted towards understanding and analyzing
the final fate of a physically realistic dynamical gravitational
collapse scenario. The current status is, despite much work in
studying the censorship and its implications, the
issue of final fate of a complete gravitational collapse
of a massive star remains far from being fully resolved.
In particular, we need to formulate in a precise
manner the conditions in gravitational collapse that
would lead to the formation of black holes
necessarily.  We now know that under
a wide variety of physically realistic situations, the
collapse ends in a black hole or a naked singularity, depending
on the initial conditions from which the collapse develops
and the dynamical evolutions as allowed by the
Einstein equations
(see e.g. \cite{Ref, Joshibook} and references therein).
It is now clear that naked
singularities are to be considered as a general feature
of general relativistic physics and that they may
develop as the end-state of collapse in a broad
variety of physical collapse situations.

It follows that a careful and extensive study of gravitational
collapse phenomena in general relativity is the key to put the theory
of black holes and their astrophysical implications on a firm footing.

From such a perspective, we investigated here
the effect of introducing small stress perturbations in the collapse
dynamics of the classic Oppenheimer-Snyder-Datt gravitational
collapse, an idealized model assuming zero
pressure, which terminates in a black hole final fate.
Our key purpose here is to study the stability
of the OSD black hole under introduction of small tangential
pressures. Clearly, stresses within a massive collapsing star are very
important physical forces to be taken into account while
considering its dynamical evolution and the final fate
of collapse (see for example
\cite{Press}).

We show here explicitly the existence of classes of
stress perturbations such that the introduction of a smallest
tangential pressure within the collapsing OSD cloud changes the
endstate of collapse to formation of a naked singularity,
rather than a black hole. It follows that the OSD
black hole is not stable under small stress perturbations
within the collapsing cloud. As we point out below,
this can also be viewed as perturbing the spacetime metric
of the cloud in a small way. Our work thus clarifies the role played
by tangential stresses in a well known gravitational
collapse scenario. The class of stress perturbations considered
here, although specific, is physically reasonable and generic
enough so as to provide a good insight into the stability
of the OSD black hole. Clearly, such a result provides an
important insight into the structure of the censorship principle
which as yet remains to be properly understood. This has
also implications towards the physical consequences
of final outcomes of a continual collapse, some of which are
indicated in the concluding remarks.

\vspace{\baselineskip}
The general spherically symmetric line element
describing the collapsing matter cloud can be written
as,
\begin{equation}\label{metric}
ds^2=-e^{2\nu(t, r)}dt^2+e^{2\psi(t, r)}dr^2+R(t, r)^2d\Omega^2,
\end{equation}
with the stress-energy tensor for a generic matter source given by,
$T_t^t=-\rho, \; T_r^r=p_r, \; T_\theta^\theta=T_\phi^\phi=p_\theta$.
The above is a general scenario, in that it involves no assumptions
on the form of the matter or the equation of state.

In order to decide on the stability or otherwise of the
OSD model under the injection of small stress perturbations, we need
to consider the dynamical development of the collapsing cloud,
as governed by the Einstein equations.
The visibility or otherwise of the final singularity is determined
by the behaviour of apparent horizon in the spacetime,
which is the boundary of the trapped surface region
that develops as the collapse progresses. First, we define a
scaling function $v(r,t)$ by the relation $R=rv$
\cite{GJ4}.
The Einstein equations for the above spacetime geometry
can then be written as,
\begin{eqnarray}\label{p2}
p_r&=&-\frac{\dot{F}}{R^2\dot{R}}, \; \rho = \frac{F'}{R^2R'} \; ,\\ \label{nu2}
\nu'&=&2\frac{p_\theta-p_r}{\rho+p_r}\frac{R'}{R}-\frac{p_r'}{\rho+p_r} \; ,\\ \label{G2}
2\dot{R}'&=&R'\frac{\dot{G}}{G}+\dot{R}\frac{H'}{H} \; ,\\
\label{F2}
F&=&R(1-G+H) \; ,
\end{eqnarray}
where the functions $F$ and $G$ are defined as,
$H =e^{-2\nu(r, v)}\dot{R}^2 , \; G=e^{-2\psi(r, v)}R'^2$.
The above are five equations in
seven unknowns, namely $\rho,\; p_r, \; p_{\theta}, \; R,\; F,\; G,\; H$.
Here $\rho$ is the mass-energy density, $p_r$ and $p_\theta$ are
the radial and tangential stresses respectively, $R$ is the physical
radius for the matter cloud, and $F$ is the Misner-Sharp mass
function.

With the above definitions of $v, H$ and $G$,  we can
substitute the unknowns $R, H$ with $v, \nu$. Without loss of
generality, the scaling function $v$ can be set $v(t_i, r)=1$ at the
initial time $t_i=0$ when the collapse commences.
It then goes to zero at the spacetime singularity $t_s$, which
corresponds to $R=0$, {\it i.e.} we have $v(t_s, r)=0$.
The above amounts to the scaling $R=r$ at the initial
epoch, which is an allowed freedom. The collapse condition
here is $\dot R<0$ throughout the evolution,
which is equivalent to $\dot{v}<0$.

We can integrate \eqref{G2} by defining a suitably regular
function $A(r, v)$ by $\nu'\equiv A_{,v}(r,v)R'$ (the function $A$ is
defined in full generality here, while often the restriction to
the class $\nu=\nu(R)$, implying $A(R)=\nu(R)$, is made, see e.g.
\cite{JGM}).
This gives $G(r,t)=b(r)e^{2rA(r,v)}$.
The arbitrary function of integration $b(r)$ can be interpreted
following the analogy with dust collapse models, where pressures vanish.
It turns out to be related to the velocity of the collapsing shells, and
once we write it as $b(r)=1+r^2b_0(r)$,
we can see that values $b_0=const.$ in the dust limit correspond
to the open ($b_0<0$), closed ($b_0>0$) or flat ($b_0=0$) Friedmann-
Robertson-Walker models. The radial stress $p_r$ and the energy
density $\rho$ are obtained from equations \eqref{p2},
once a specific choice for the mass function $F(r,t)$ is made. The function
$\nu$ can be taken as the second free function for the system so
that once a particular form of $\nu$ is specified, equation \eqref{nu2}
provides the tangential stress profile $p_{\theta}$. Finally, from the
equation of motion \eqref{F2}, we can integrate to obtain $v(r, t)$,
thus solving the system of Einstein equations.

We can also invert the function $v(r,t)$,
which is monotonically decreasing in $t$, to obtain the
time needed by the matter shell at any radial value $r$ to
reach the event with a particular value $v$. We write
the function $t(r, v)$ from equation \eqref{F2} as,
\begin{equation}
t(r, v)= \int^1_{v}\frac{e^{-\nu}}{\sqrt{\frac{F}{r^3\tilde{v}}+\frac{be^{2rA}-1}{r^2}}}d\tilde{v} \; .
\end{equation}
The time taken by the shell at $r$ to reach the spacetime singularity
at $v=0$ is then $t_s(r)=t(r, 0)$.

Since $t(r, v)$ is in general at least $C^2$ everywhere in the spacetime
(because of the regularity of the functions involved), and is continuous at the
center, we can write it as,
\begin{equation}\label{t}
t(r, v)= t(0, v)+r\chi(v)+O(r^2) \;
\end{equation}
When $t(r,v)$ is differentiable, we can make a Taylor expansion
near the center $r=0$.
Here, $t(0,v)$ is the above integral evaluated at $r=0$
and $\chi(v)=\left.\frac{dt}{dr}\right|_{r=0}$. As we point out below, the
quantity $\chi(0)$ plays an important role towards determining the
nature of the final singularity of collapse.
We consider collapse from a regular initial data, and
so the Einstein equation (5) implies that the Misner-Sharp
mass $F(r, v)$
must go as $r^3$ near the center $r=0$ in order for the density to be regular
at the center, and also to have $t(0, v)$ well defined.
Therefore, in general, $F$ must have the form,
$F(r,v) = r^3 M(r,v)$,
where $M$ is a suitably regular function.
Then, by continuity, the time for the shell located at any $r$ close to
center to reach the singularity is given as,
$t_s(r)= t_s(0)+r\chi(0)+O(r^2)$.
Basically, this means that the singularity curve should have a well-defined
tangent at the center. Regularity at the center also implies that
the metric function $\nu$ cannot have constant or linear terms in
$r$ in a close neighborhood of $r=0$, and it must go as  $\nu\sim r^2$
near the center. Therefore the most general choice of the free function
$\nu$ is,
\begin{equation}
\nu(r,v)=r^2g(r,v) \;
\end{equation}
Since $g(r, v)$ is a regular function (at least $C^2$), it can be
written near $r=0$ as,
\begin{equation}\label{expand-g}
g(r, v)=g_0(v)+g_1(v)r+g_2(v)r^2+...
\end{equation}

We would now like to investigate how the OSD gravitational
collapse scenario, which is a homogeneous pressureless dust cloud collapsing
to give rise to a black hole, gets altered when small stress perturbations are
introduced in the dynamical evolution of collapse.

To that end, we first note that the dust scenario is obtained if
$p_r=p_{\theta}=0$ in the above. In that case, from equation \eqref{nu2} it
follows that $\nu'=0$ and that together with the condition $\nu(0)=0$ gives
$\nu=0$ identically.
These models have been widely studied in the literature, and it is seen
that for generic dust collapse the final outcome can be either a black hole
or a naked singularity, depending on the nature of the initial density
and velocity profiles of the collapsing matter shells
\cite{dust}.
In the OSD collapse to a black hole, the trapped surfaces
or the apparent horizon in the spacetime develop much earlier before
the formation of the final singularity of collapse. On the other hand,
when inhomogeneities are allowed in the initial density profile, such as a
higher density at the center of the star, then the trapped surface formation
is delayed in a natural manner within the collapsing cloud and the final
singularity becomes visible to faraway observers in the universe
\cite{JDM}.

The OSD case is obtained from above when
we further assume that the collapsing dust is necessarily homogeneous at all
epochs of collapse. This is of course an idealized scenario because realistic
stars would have typically higher densities at the center, which slowly falls
off with increasing radius, and they also would have non-zero internal stresses.
Specifically, the conditions that must be imposed to obtain the OSD case
from the above are given by,
\begin{itemize}
  \item[(a)] $M=M_0$\; ,
  \item[(b)] $v=v(t)$\; ,
  \item[(c)] $b_0(r)=k$\; .
\end{itemize}
Then we have $F'=3M_0r^2$, $R'=v$, and the energy density is homogeneous
throughout the collapse with $\rho=\rho (t)= {3M_0}/{v^3}$.
The spacetime geometry then becomes the Oppenheimer-Snyder metric,
\begin{equation}
ds^2=-dt^2+\frac{v^2}{1+kr^2}dr^2+r^2v^2d\Omega^2 \; ,
\end{equation}
where the function $v(t)$ is solution of the equation of motion,
$ \frac{dv}{dt}=\sqrt{(M_0/v)+k}$,
obtained from Einstein equation \eqref{F2}. In this case we get
$\chi(0)=0$ identically. All the matter shells then collapse into a
simultaneous singularity (due to condition (b)), which
is necessarily covered by the event horizon that developed in the
spacetime at an earlier time, thus giving rise to a black hole.

To examine the effect of introducing stress perturbations
in the above scenario and to study the models thus obtained which are close
to the Oppenheimer-Snyder  in this sense, we need to relax
and perturb one or more of the above conditions (a), (b) or (c).

If the collapse outcome would not to be a black hole, the final
singularity of collapse cannot be simultaneous.  We are thus led to relax
condition (b) above, allowing $v = v(t,r)$, rather than $v=v(t)$ only.
At the same time, in order not to depart too much from the OSD model,
we keep (a) and (c) unchanged. This also brings out more clearly the role
played by the stress perturbations in the model.

In terms of the spacetime metric \eqref{metric},
while the metric function $\nu(t,r)$ must be identically vanishing for the
dust case, the above amounts to allowing for small perturbations in $\nu$,
and allowing it to be non-zero now. This is equivalent to introducing
small stress perturbations in the model, and we show below how that
affects the apparent horizon developing in the collapsing cloud.

We note immediately that taking $M=M_0$ leads to $ F=r^3M_0$.
We have $R'=v+rv'\rightarrow v$
for $r\rightarrow 0$ and therefore we get
$ A_{,v}= {\nu'}/{v}$.
With the expansion near $r=0$ for
both $A$ and $g$ we get the relation between the coefficients of
the expansion of $g$ and those
for the expansion of $A$. Integrating \eqref{G2} in the small $r$ limit we thus
obtain $G(r,t)=b(r)e^{2\nu(r, v)}$.
The radial stress $p_r$ vanishes in this case as $\dot F=0$, while
the tangential pressure, obtained from equation \eqref{nu2}, has the form,
$p_\theta= p_1 r^2 + p_2 r^3 +...$, where $p_1, p_2$ are naturally
evaluated in terms of coefficients of $m, g$, and $R$ and its derivatives,
\begin{equation}\label{pt}
p_\theta=3\frac{M_0g_0}{vR'^2}r^2+\frac{9}{2}\frac{M_0g_1}{vR'^2}r^3+...
\end{equation}
Here the choice of sign of the functions $g_0$
and $g_1$ is enough to ensure positivity or negativity of $p_\theta$.

We note that scenarios with vanishing radial stresses but
non-vanishing tangential stresses have been considered in past,
with the most physically significant model (though not the only relevant one)
being the so called `Einstein cluster' (see
\cite{cluster}),
which describes a cloud of collapsing counter rotating particles.
Naked singularities and black holes are found to arise as the endstate of
such models, depending on the initial density, velocity and stress configurations \cite{GJ5}.

The first order coefficient $\chi$ in
equation of the time curve of the singularity $t_s(r)$ is now obtained as
\begin{equation}\label{chi}
\chi(0)=-\int^1_0\frac{v^{\frac{3}{2}}g_1(v)}{(M_0+vk+2vg_0(v))^{\frac{3}{2}}}dv \; .
\end{equation}
As mentioned above, it is $\chi(0)$ that governs the
nature of the singularity curve, and whether it is increasing or decreasing
away from the center. Clearly, it is the matter initial data in terms of
density and stress profiles, the velocity of the collapsing shells, and
the allowed dynamical evolutions that govern and fix
the value of $\chi(0)$.

The quantity $\chi(0)$ also governs the behaviour of
apparent horizon and trapped surface formation, as we show
below, which in turn governs the nakedness or otherwise of
the singularity. The equation for the apparent horizon is
given by ${F}/{R}=1$. It is analogous to that of the dust case
since ${F}/{R}={rM}/{v}$ in both cases
\cite{JDM}.
So the apparent horizon curve $r_{ah}(t)$ is given by
$ r_{ah}^2=\frac{v_{ah}}{M_0}$,
with $v_{ah}=v(r_{ah}(t), t)$, which can also be inverted as a
time curve $t_{ah}(r)$.
The visibility of the singularity at the center of the collapsing cloud
to faraway observers is determined by the
nature of this apparent horizon curve which is given by,
\begin{equation}\label{t-ah}
t_{ah}(r)=t_s(r)-\int_0^{v_{ah}}\frac{e^{-\nu}}{\sqrt{\frac{M_0}{v}+\frac{be^{2\nu}-1}{r^2}}}dv
\end{equation}
where $t_s(r)$ is the singularity time curve, whose initial point is
$t_0=t_s(0)$. Near $r=0$ the equation \eqref{t-ah} becomes,
\begin{equation}
    t_{ah}(r) =t_0+\chi(0)r+o(r^2) \; .
\end{equation}

From the above, it is now easy to see how the stress perturbation
affects the time of formation of the apparent horizon, and therefore the
formation of a black hole or naked singularity.
A naked singularity typically occurs as collapse endstate when a
co-moving observer at fixed $r$ does not encounter any trapped surfaces
till the time of singularity formation. For a black hole to form, trapped
surfaces develop before the singularity, so it is needed that,
\begin{equation}
t_{ah}(r) \le t_0 ~~\mbox{for}~~ r>0, ~~\mbox{near}~~ r=0 \; .
\end{equation}
It is clear that for all functions $g_1(v)$ for which $\chi(0)$ is
positive, this
condition is violated and the apparent horizon is forced to appear
after the formation of the central singularity. The apparent horizon curve
then initiates at the central singularity $r=0$ at $t=t_0$ and increases
with increasing $r$, moving to the future, {\it i.e.} $t_{ah} > t_0$ for
$r > 0 $ near the center. The behaviour of outgoing families of
null geodesics has been analyzed in detail in such a case
when $\chi(0)>0$ and we know
that geodesics terminate at the singularity in the past.
Thus timelike and null geodesics come out from the singularity,
making it visible to external observers
\cite{Geo}.

It follows that $g_1$ is the term in the stresses $p_\theta$
which decides the black hole or naked singularity final fate.
We can choose it to be arbitrarily small, and we now see how introducing
a generic tangential stress perturbation in the model would change
drastically the final outcome of collapse. For all non-vanishing
tangential stresses with $g_0=0$ and $g_1<0$, even
the slightest perturbation of the Oppenheimer-Snyder-Datt
scenario, injecting a small tangential stress would result in a naked
singularity. The space of all functions $g_1$ that make $\chi(0)$
positive, which includes all the strictly negative functions $g_1$,
causes the collapse to end in a naked singularity. We note that
while this is an explicit example, by no means this is the only
class.

The remarkable feature of this class is that it
corresponds to a collapse model for a simple and straightforward
perturbation of the Oppenheimer-Snyder-Datt spacetime metric,
where the geometry near the center can be written as,
\begin{equation}\label{pert}
ds^2=-(1-2g_1 r^3)dt^2+\frac{(v+rv')^2}{1+kr^2-2g_1 r^3}dr^2+r^2v^2d\Omega^2 \; ,
\end{equation}
The metric above satisfies Einstein equations in the neighborhood
of the center of the cloud when the function $g_1(v)$ is small and bounded.
We could take for example,  $0<|g_1(v)|<\epsilon$, so that the smaller we
take the parameter $\epsilon$ the bigger will be the radius where the
approximation is valid.
The function $v(r,t)$ above is governed by the equation of motion
\eqref{F2} which in the small $r$ limit becomes,
$ {dv}/{dt}=(1-g_1(v) r^3) ({\frac{M_0}{v}+k-2g_1(v) r})^{1/2}.$
Finally, $\chi(0)$ in this case is given by equation \eqref{chi} with $g_0=0$,
and  in certain cases can also be integrated.


We note that any realistic matter model must satisfy some
energy conditions ensuring the positivity of mass and energy density.
In general, the weak energy condition implies restrictions on the density
and pressure profiles. The energy density as given by the second of equations
\eqref{p2} must be positive. Since $R$ is positive, to ensure positivity
of $\rho$ we require $F>0$ and $R'>0$. The choice of
positive $M(r)$ (which obviously holds for $M_0>0$ and is physically reasonable)
ensures positivity of the mass function. Then $R'>0$ is a sufficient condition for the
avoidance of shell crossing singularities. The tangential stress can be written
from \eqref{nu2} where $p_r=0$, and is given by $p_\theta=\frac{1}{2}\frac{R}{R'}\rho\nu'$.
So the sign of the function $\nu'$ determines the sign of $p_\theta$.
Positivity of $\rho+p_\theta$ is then ensured for small values of $r$
throughout collapse for any form of $p_\theta$. In fact, regardless
of the values taken by $M$ and $g$, there will always be a neighborhood of $r=0$
for which $|p_\theta|<\rho$ and therefore $\rho+p_\theta\geq0$.

%

The black hole and naked singularity outcomes of gravitational
collapse are very different from each other physically, and would have
quite different observational signatures. In the naked singularity case we have
the possibility to observe the physical effects happening in the vicinity of
the ultra dense regions that form in the very final stages of collapse.
However, in a black hole scenario, such regions are necessarily hidden
within the event horizon. The fact that a slightest stress perturbation
of the OSD collapse could change the outcome drastically, taking it from a
black hole to naked singularity formation, means that the naked singularity
final state for a collapsing star must be studied carefully to deduce its
physical consequences which are not well understood so far.

The existence of subspaces of collapse solutions as we have
shown here, that go to a naked singularity final state rather than a black
hole, in the arbitrary vicinity of the OSD black hole, presents an intriguing
scenario. It gives an idea of the richness of the structure present in
gravitation theory and the complex solution space of Einstein equations
which are a complicated set of highly non-linear partial differential equations.
What we see here is there are classes of stress perturbations such
that an arbitrarily small change from the OSD model is a solution going
to naked singularity. In this sense, this manifests an instability in the
black hole formation process in gravitational collapse. This also provides
an intriguing insight into the nature of cosmic censorship, namely that
the collapse must be properly fine-tuned necessarily if it is to produce
a black hole only as the final endstate.

Traditionally it was believed that the presence of
stresses or pressures in the collapsing matter cloud would increase the chance
of black hole formation, thereby ruling out dust models
that were found to lead to a naked singularity as collapse endstate.
That is no longer the case.
The model described here not only provides a new class of collapses
ending in a naked singularity, but more importantly, shows how the bifurcation
line that separates the phase space of `black hole formation' from that of
the `naked singularity formation' runs directly over the simplest and most
studied of black hole scenarios such as the OSD model, thus making
it unstable under perturbations.

\end{document}